\documentclass[prd,superscriptaddress,footinbib,twocolumn,floatfix]{revtex4}
\usepackage{amssymb}

\makeatletter
\renewcommand{\@makecaption}[2]{
  \vskip\abovecaptionskip
  \sbox\@tempboxa{\small\sf #1: #2}%
  \ifdim \wd\@tempboxa >\hsize
  \small\sf #1: #2\par
  \else
    \global \@minipagefalse
    \hb@xt@\hsize{\hfil\box\@tempboxa\hfil}%
  \fi
  \vskip\belowcaptionskip}
\makeatother

\def\ba{\begin{eqnarray}}
\def\ea{\end{eqnarray}}

\newcommand{\nn}{\nonumber}

\newcommand{\ft}[2]{{\textstyle\frac{#1}{#2}}}
\newcommand{\eqn}[1]{(\ref{#1})}

\def\Dslash{\,\,{\raise.15ex\hbox{/}\mkern-12mu D}}
\def\Dbarslash{\,\,{\raise.15ex\hbox{/}\mkern-12mu {\bar D}}}
\def\delslash{\,\,{\raise.15ex\hbox{/}\mkern-9mu \partial}}
\def\delbarslash{\,\,{\raise.15ex\hbox{/}\mkern-9mu {\bar\partial}}}
\def\pslash{\,\,{\raise.15ex\hbox{/}\mkern-9mu p}}
\def\calDslash{\,\,{\raise.15ex\hbox{/}\mkern-12mu {\cal D}}}

\begin{document}
\preprint{HUTP-01/A067}
\preprint{MIT-CTP-3247}

\title{D-Brane Probes of $G_2$ Holonomy Manifolds}

\author{Sergei \surname{Gukov}}
\email{gukov@gauss.harvard.edu}
\affiliation{Jefferson Physical Laboratory,
  Harvard University,
  Cambridge MA 02138}
\author{David Tong}
\email{dtong@mit.edu}
\affiliation{Center for Theoretical Physics,  
Massachusetts Institute of Technology, 
Cambridge, MA 02139}

\begin{abstract}
We describe how mirror symmetry of three-dimensional ${\cal N}=1$ 
supersymmetric gauge theories can be used to determine the 
theory on the world-volume of a D2-brane probe of manifolds with 
$G_2$-holonomy. This is a much shortened companion paper to 
\cite{gt}.
\end{abstract}


\maketitle

\newcommand{\be}{\begin{equation}}
\newcommand{\ee}{\end{equation}}

\subsection*{Introduction}

The concept of the D-brane probe provides the link between gauge 
theory and geometry. It has proven to be one of the most 
important ideas among recent developments in string theory, 
shedding new light on strong coupling gauge 
theory dualities, providing a physical interpretation of 
previously obscure algebraic-geometric constructions, and 
resulting in examples of the AdS/CFT correspondence with 
increasingly realistic gauge and matter content. 

The purpose of this letter is to describe some results 
extending this enterprise to D2-brane probes of non-compact 
manifolds of $G_2$ holonomy. 
These spaces are of interest for both theoretical and 
phenomenological reasons \cite{bob,amv,aw,bobed,decon}. 
For orbifold backgrounds, and their partial resolutions,
there are well known procedures for determining the gauge theory
living on the probe world-volume \cite{dm,drm}. (See 
\cite{tsimpis} for examples of $G_2$ holonomy orbifolds). 
However, recent attention has focused on
asymptotically conical $G_2$ manifolds for which the orbifold 
prescription is not relevant.
Here we suggest a different approach to understand 
aspects of the probe theory,
using mirror symmetry of three dimensional gauge theories \cite{is}.

Recall that mirror symmetry is a duality between a pair of 
three-dimensional field theories which, among other things, 
interchanges the Coulomb and Higgs branches. The basic observation 
is that mirror symmetry on an M2-brane probe has a simple 
interpretation as an ``M-theory flip'' \cite{pz}:
$$
\matrix{
& & \mbox{M-theory on $X$} & & \cr
& \swarrow & & \searrow & \cr
\mbox{IIA with} & & & & \cr
\mbox{D6-branes} && &&  \mbox{IIA on $X$}\cr
\wr & & & & \wr \cr
\mbox{Coulomb}& & \mbox{mirror symmetry} & & \mbox{Higgs} \cr
\mbox{branch}& & \Longleftrightarrow & & \mbox{branch}}
$$
In order to explain this duality, let us consider a membrane probe
on a (singular) space $X$ with $G_2$ holonomy.
We can perform a reduction from M-theory to IIA string theory
in two different ways.
First, we can reduce on a circle transverse to both the manifold $X$, 
and to the membrane. This takes us to IIA theory on $X$.
The space $X$ is reproduced as a Higgs branch in the D2-brane
world-volume theory, in a way reminiscent to the hyperK\"ahler quotient
construction. By analogy, we call it the $G_2$ quotient construction.

On the other hand, we can reduce on an ${\bf S}^1$ contained within $X$.
There are many ways to choose the ${\bf S}^1$, which lead to
different IIA backgrounds with D6-branes and/or Ramond-Ramond fluxes.
An illustrative example was described in \cite{amv}.
One might ask: ``What is the natural choice of the M-theory circle?''. 
One simple choice is to require $X/U(1) \cong \mathbb{R}^6$.
If such a quotient exists it gives, after reduction to type IIA 
theory, a configuration of D6-branes in a (topologically) flat 
space-time. The positions of the D6-branes correspond to the fixed 
points of the circle action. In contrast to the previous reduction,
the geometry of $X$ is entirely encoded in the configuration of 
D6-branes, rather than in the geometry of space-time \cite{aw,GS}. 
The M-theory geometry $X$ is then re-constructed as the Coulomb branch
of the world-volume theory on the D2-brane.
When the space $X$ develops a conical singularity,
the configuration of D6-branes also becomes singular.
In particular, in some cases of interest it degenerates into 
a collection of flat D6-branes intersecting at the suitable angles 
\cite{BDL}. For such models, it becomes a simple exercise to derive
the three-dimensional gauge theory on the D2-brane probe. Then, using 
mirror symmetry, one can obtain the theory on the probe of $X$.

The simplest illustration of this method was given in \cite{pz},
where a single D2-brane probe of $N$ parallel D6-branes was used
to re-derive the original $\mathcal{N}=4$ mirror pairs of Intriligator 
and Seiberg \cite{is}. Note, however, that we here we run the logic 
of \cite{pz} in reverse: we use mirror symmetry, derived through 
independent techniques, to derive the gauge theory on the 
probe of $X$.

This letter contains only the barest details of our method. 
Many further results and examples, including the derivation of 
the mirror pairs and the interesting subtleties involved,
as well as applications to $SU(3)$, $Sp(2)$ and $Spin(7)$ 
holonomy manifolds can be found in \cite{gt}.

\subsection*{Mirror Symmetry}

Let us firstly describe the mirror pairs that will be 
our tool in understanding the probe theories. Our 
mirror pairs have ${\cal N}=1$ supersymmetry, and  
are derived in \cite{gt} from deforming ${\cal N}=4$ 
mirror pairs using both field theory and string theory 
techniques. Of course with such little supersymmetry 
(${\cal N}=1$ means 2 supercharges) we have little control 
over the strong coupling dynamics and must be wary of any 
conjectured duality. 
Our only savior is parity symmetry 
which may be used to prohibit the lifting of certain 
vacuum moduli spaces \cite{ahw,GK}. 
We hope that the success of our mirror pairs in describing
manifolds of $G_2$ holonomy goes some way towards convincing
the reader of their utility. 

The mirror pairs preserve only ${\cal N}=1$ supersymmetry, 
and are given by
\begin{eqnarray}
{\rm \bf Theory\ A:} && \ U(1)^r\ \mbox{with $k$ scalars and $N$ hypers}
\nn\\
{\rm \bf Theory\ B:} && \ U(1)^{N-r}\ \mbox{with $(3N-k)$ 
scalars and} \nn\\ &&\ \mbox{\ $N$ hypermultiplets} \nn
\end{eqnarray}
The abelian vector multiplets contain only a photon and 
a Majorana spinor, while the scalar multiplets, which we 
shall denote as $\Phi$, contain a single real scalar
and a Majorana fermion.
In contrast, the hypermultiplets fill out representations of 
the ${\cal N}=4$ algebra: they each contain four Majorana 
fermions and two complex scalars, $q$ and $\tilde{q}$. 
We write the superfield as a doublet, $W=(Q,\tilde{Q}^\dagger)^T$.
The chiral multiplets $Q$ and $\tilde{Q}$ carry conjugate charges 
under the gauge group. For Theory A, we denote the charge of 
the hypermultiplets as $R_i^a$, while for Theory B it  
is $\hat{R}_i^p$; $i=1,\ldots,N\ ;a=1,\ldots,r;\ 
\ p=1,\ldots, N-r$. Each of these matrices are assumed to 
be of maximal rank, and mirror symmetry requires
\begin{equation}
\sum_{i=1}^NR_i^a\hat{R}_i^p=0\ \ \ \ \ \ \ \ \forall\ a,p
\end{equation}
In ${\cal N}=1$ theories, there are no holomorphic 
luxuries and interactions are determined in terms 
of a {\it real} superpotential, $f$. For Theory A, 
this superpotential has the cubic form associated with 
the ${\cal N}=4$ theories, and is determined by a 
triplet of $k\times N$ matrices ${T}_c$, $c=1,2,3$
\begin{equation}
f=\sum_{i=1}^N\sum_{c=1}^3\sum_{\alpha=1}^k 
W_i^\dagger\, {\tau}^c\,W_i\cdot
{T}_{c,i}^\alpha\,\Phi_\alpha
\label{super}\end{equation}
where $\alpha=1,\ldots,k$ and ${\tau}^c$ are the three 
Pauli matrices. A similar coupling exists for Theory B, now 
with the triplet of $(3N-k)\times N$ coupling matrices 
$\hat{T}_c$, satisfying
\begin{equation}
\sum_{i=1}^N\sum_{c=1}^3{T}_{c,i}^\alpha\,
{\hat{T}}_{c,i}^\rho=0
\ \ \ \ \ \forall\ \alpha,\rho
\label{tt}\end{equation}
Further details of these theories, together with the 
methods used to derive them, can be found in \cite{gt}. 
Here let us restrict ourselves to a few comments. The 
Coulomb branch of Theory A  has dimension $(N+r)$,
which coincides with the dimension of the 
Higgs branch of Theory B. (The converse also holds). 
Mass and FI parameters may be added to both 
theories, partially lifting some branches of vacua, and 
the mirror map for these deformations is known. 

\subsection*{The $G_2$ Quotient Construction}

Let us now apply the mirror pairs described above to a 
D2-brane probe of a D6-brane background. We take 
$i=1,\ldots,N$, flat D6-branes, each of which has spatial 
world-volume direction,
\begin{eqnarray}
D6_i:&&\ \  123[47]_{\theta_1^i}[58]_{\theta_2^i}
[69]_{\theta_3^i}
\nn\end{eqnarray}
The D6-branes lie on a special lagrangian locus if each 
rotation is contained in $SU(3)$ \cite{BDL} or, more simply, if
\begin{equation}
\theta^i_1\pm\theta_2^i\pm\theta_3^i=0\ \ \ \ \ {\rm mod}\ 2\pi\ 
\forall\ i
\label{lorientation}\end{equation}
ensuring that ${\cal N}=1$ supersymmetry (4 supercharges) is 
preserved on their common world-volume. (For non-generic 
angles, more supersymmetry may be preserved. We will assume 
this is not the case). 

As described in the introduction, we probe this configuration with a 
D2-brane lying in the $x^1-x^2$ plane. 
This breaks supersymmetry by a further half, resulting in a $d=2+1$ 
dimensional world-volume theory with ${\cal N}=1$ supersymmetry 
(2 supercharges). For the 
singular case of intersecting, flat D6-branes, the theory on 
the D2-brane probe is simple to write down. The 2-2 strings give 
rise to the usual gauge field and seven scalars. Of these, there is 
one free ${\cal N}=1$ scalar multiplet parameterizing motion
in the $x^3$ direction common to all D6-branes. Further fields 
arise from the 2-6 strings. These give rise to $N$ hypermultiplets. 
Thus, we have the interacting ${\cal N}=1$ supersymmetric
theory on the probe, 
\begin{center}
{\bf Theory A:} $U(1)$ with 6 scalar multiplets and N hypermultiplets
\end{center}
where each hypermultiplet has charge $+1$ under the gauge field. 
The couplings of the hypermultiplets to the scalar multiplets 
are determined by the geometry of the D6-branes: each 
hypermultiplet couples minimally to the three scalar fields 
orthogonal to the corresponding D6-brane. If we define the scalar 
fields $\phi_\alpha=x^{\alpha+3}$, $\alpha=1,\ldots,6$,
then the superpotential is of the form \eqn{super} with the 
couplings determined by the D6-brane orientations,
\begin{equation}
{T}^\alpha_{c,i}=-\sin\theta_c^i\,\phi_\alpha\,\delta_{c,\alpha} 
+\cos\theta_c^i\,\phi_{\alpha}\,\delta_{c,\alpha-3}\ \ \ \ \ \ c=1,2,3
\label{ttheta}\end{equation}
{}From the IIA space-time picture, we are lead to the natural conjecture 
that the Coulomb 
branch of this theory, parameterized by the six real scalars 
$\phi_\alpha$, together 
with the dual photon $\sigma$, is a seven dimensional manifold $X$ that 
admits a metric of $G_2$ holonomy. However, this description of $X$ in terms 
of Coulomb branch variables is not overly useful. In particular, the 
isometries of $X$ are lost in the reduction to IIA, and are only expected to 
be recovered as isometries of the Coulomb branch in the strong coupling limit. 
It would be desirable to have an algebraic description of $X$, in which 
the symmetries are manifest.
This is exactly what the mirror theory provides for us. 

Since Theory A is in the class of theories discussed above, we 
may simply write down the mirror theory whose Higgs 
branch is conjectured to give the $G_2$ manifold $X$,
\begin{center}
{\bf Theory B:} $U(1)^{N-1}$ with $3(N-2)$ scalar and $N$ 
hypermultiplets 
\end{center}
The gauge couplings are determined by the $A_{N-1}$ quiver diagram: i.e 
the $i^{\rm th}$ gauge group acts on the $i^{\rm th}$ 
hypermultiplet with charge $+1$, and the $(i+1)^{\rm th}$ 
hypermultiplet with charge $-1$. All other hypermultiplets are 
neutral. The Yukawa terms are of the form \eqn{super}, 
with the triplet of coupling matrices $\hat{T}$ determined by \eqn{tt}. 
The Higgs branch of this theory is parameterized by 
$w_i=(q_i,\tilde{q}_i^\dagger)^T$, the $N$ doublets of complex scalars 
in the hypermultiplets. These are constrained by the 
$3(N-2)$ D-terms, modulo $(N-1)$ $U(1)$ gauge quotients,
\begin{equation}
\sum_{i,c}\hat{T}^{\rho}_{c,i} w_i^\dagger\tau^c w_i = 0\ \ \ \ \ \ 
\rho=1,\ldots,3(N-2)
\label{g2c}\end{equation}
This quotient construction yields a conical manifold,
which is expected to admit a metric of $G_2$ holonomy.
In some cases the conical singularity may be (partially) resolved
by adding constants to the right-hand side of \eqn{g2c}.
This blows up two-cycles and, in the IIA picture, corresponds to 
translating the D6-branes in the $x^4-x^9$ directions. 
Note that when the Yukawa 
matrices $\hat{T}$ fall into suitable $SU(2)$ triplets, 
the above method coincides with the toric hyperK\"ahler 
quotient construction, supplemented by a further quotient by a  
tri-holomorphic isometry to yield a manifold of dimension 
seven. This is the construction discussed by Acharya and 
Witten \cite{bobed}. However, in general, the charges in \eqn{g2c} 
differ. 

The above theory may also be considered as a ${\cal N}=(1,1)$ 
supersymmetric linear sigma model in $d=1+1$ dimensions, {\it cf.} \cite{av}.
However, in the absence of something akin to Yau's theorem,
we cannot be sure that the Ricci flat metric to which
the theory flows has $G_2$ holonomy. 

\subsection*{An Example}

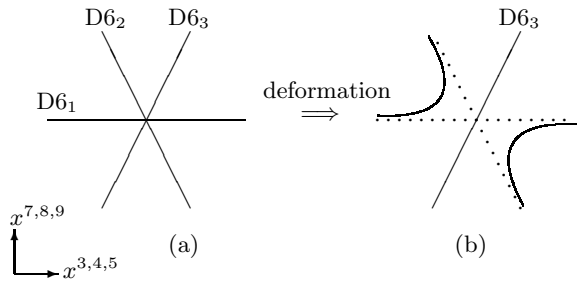
\begin{figure}

\setlength{\unitlength}{0.9em}
\begin{center}
\begin{picture}(22,11)

\put(7,2){\line(-1,2){4}}\put(2.2,10.3){D6$_2$}
\put(3,2){\line(1,2){4}}\put(6,10.3){D6$_3$}
\put(0.5,6){\line(1,0){9}}\put(0,6.5){D6$_1$}

\put(12,6){$\Longrightarrow$}
\put(10.3,7){deformation}

\multiput(22,2)(-0.25,0.5){16}{\circle*{.1}}
\put(18,2){\line(1,2){4}}\put(21,10.3){D6$_3$}
\multiput(15.5,6)(0.5,0){18}{\circle*{.1}}
\qbezier(15.5,6.2)(20,6)(17.8,9.8)
\qbezier(22.1,2.1)(20,6)(24.5,5.8)

\put(6,0){(a)}\put(19,0){(b)}


\put(-1,-1){\vector(1,0){2}}\put(-1,-1){\vector(0,1){2}}
\put(1.2,-1.2){$x^{3,4,5}$}\put(-1.2,1.2){$x^{7,8,9}$}

\end{picture}\end{center}
\caption{Intersection of special Lagrangian D6-branes dual
to M-theory on $G_2$ holonomy cone over $SU(3)/U(1)^2$ (a),
and its non-singular deformation (b).}
\label{fig}
\end{figure}

Let us now examine the $G_2$-quotient construction applied to 
a specific example. Our choice for consideration is the 
$G_2$ manifold $X$ given by the cone 
over the flag manifold $SU(3)/U(1)^2$ \cite{BS,gary}.
This example was also discussed in detail by Atiyah and Witten \cite{aw}.
They show that, with a suitable choice of M-theory circle, 
$X$ can be reduced to three, flat, intersecting D6-branes
in type IIA string theory.
The symmetry of $X$ (to be discussed below), together with the 
special lagrangian condition \eqn{lorientation} determines 
the angles of these three branes to be $\theta^1_c=0$, 
$\theta^2_c={2\pi}/{3}$ and $\theta_c^3={4\pi}/{3}$ for each $c$. 
The configuration is drawn in Figure \ref{fig}.

In order to make the symmetries of the configuration manifest,
we define two triplets of scalars, $\vec{\phi}_1=(x^7,x^8,x^9)^T$
and $\vec{\phi}_2=(x^4,x^5,x^6)^T$ in terms of which the orientation 
of the $i^{\rm th}$ D6-brane can be described by the set of linear equations. 
\begin{eqnarray}
D6_1 :&&\ \ \vec{\phi}_1=0 \nn\\
D6_2 :&&\ \ \ft12\vec{\phi}_1+\ft{\sqrt{3}}{2}\vec{\phi}_2=0 \nn\\
D6_3 :&&\ -\ft12\vec{\phi}_1+\ft{\sqrt{3}}{2}\vec{\phi}_2=0
\nn\end{eqnarray}
The original $G_2$ holonomy manifold $X$ enjoyed an $SU(3)$ 
continuous isometry. It's not surprising that, upon taking 
the quotient to IIA string theory, this isometry is partially lost.
In fact, the D6-brane background has only a $SU(2)$ symmetry,
under which each $\vec{\phi}_a$ transforms as a triplet.
The M-theory circle itself provides one further, hidden, 
$U(1)$ action. We therefore conclude that the reduction to IIA string 
theory has broken the isometry group to 
$SU(3)\rightarrow SU(2)\times U(1)$. 
Now, let us introduce a probe D2-brane in this background,
and look at the $\mathcal{N}=1$ gauge theory on its world-volume.
\begin{center}
{\bf Theory A:} $U(1)$ with 6 scalars and 3 hypermultiplets
\end{center}
As described above, the 6 scalar multiplets combine into two triplets 
whose interactions with the hypermultiplets are of the form \eqn{super}, 
where the interaction matrices are determined by \eqn{ttheta}. 
The Coulomb branch of this theory is parameterized by the six 
scalars, together with the dual photon. It has the 
$SU(2)\times U(1)$ isometry group, which is 
expected to be enhanced to the full $SU(3)$ only in the 
strong coupling, infra-red limit. 

Using the results described earlier, the mirror theory is the 
${\cal N}=1$ gauge theory with matter content,
\begin{center}
{\bf Theory B:} $U(1)^2$ with 3 scalar and 3 hypermultiplets
\end{center}
The charges of the three hypermultiplets under the $U(1)^2$ gauge group 
are $(+1,-1,0)$ and $(0,+1,-1)$. The three scalars couple 
through the usual superpotential \eqn{super}, with interactions 
determined by \eqn{tt} and \eqn{ttheta} to be 
$\hat{T}_{c,i}^\rho=\delta^\rho_c$ for all $i$. 
Let us examine the Higgs branch of this theory.
The superpotential provides 3 
real constraints on the 12 real scalar fields contained 
in the hypermultiplets. After dividing by the gauge group, we are 
left with a Higgs branch of dimension 7, as required. The 
constraints are,
\begin{eqnarray}
\sum_{i=1}^3|q_i|^2-|\tilde{q}_i|^2=0,\quad\quad
\sum_{i=1}^3\tilde{q}_iq_i=0
\label{itsnice}
\end{eqnarray}
Firstly notice that this space has a manifest $SU(3)$ 
symmetry, thus recovering the full isometry group of $X$. It 
is not difficult to further show that the space is indeed 
isomorphic to the cone over $SU(3)$, ensuring that the 
full Higgs branch is the flag manifold $SU(3)/U(1)^2$. 

There is a single normalizable deformation of this space,
which yields a smooth $G_2$ manifold:
\begin{equation}
X\cong \mathbb{R}^3\times \mathbb{C}\bf{P}^2
\nn\end{equation}
In the D6-brane picture, the singularity is resolved by deforming 
the singular locus of flat, intersecting D6-branes into a smooth special 
Lagrangian curve $L \subset \mathbb{C}^3$:
\begin{equation}
L\cong\mathbb{R}\times \bf{S}^2 \cup \mathbb{R}^3
\nn\end{equation}
In the present case this deformation 
involves only two out of the three D6-branes.
To see this more explicitly, let us choose the first
and second D6-branes, which deform to lie 
on the special lagrangian curve:
\begin{equation}
\vec{\phi}_1\cdot\vec{\phi}_2=-|\vec{\phi}_1||\vec{\phi}_2|, 
\quad\quad |\vec{\phi}_1|\cdot(3|\vec{\phi}_1|^2-|\vec{\phi}_2|^2)=\rho
\label{slag}\end{equation}
This curve has a remarkable property: it creates a hole through 
which the remaining D6-brane can pass, see Figure \ref{fig}.
Therefore, it suffices to 
deform only two of the three D6-branes in order to completely 
remove the conical singularity. Of course, one has three different 
ways to pick a pair of D6-branes, leading to three different 
resolutions of the space, meeting at a singular point. This is 
precisely the picture suggested in \cite{aw}.

It is natural to ask how the probe theory responds to such a deformation. 
{}From the perspective of Theory A, one can show that there is essentially 
a unique deformation consistent with all the symmetries of the model; 
it is a Yukawa term coupling a pair of hypermultiplets. Moreover, the 
locus of zeroes of the fermion mass matrix has the same topology as 
the locus \eqn{slag}. For more details, see \cite{gt}.

\section*{Acknowledgements}

We are grateful to
B.~Acharya, M.~Aganagic, N.~Constable, A.~Hanany, 
J.~Sparks, N.~Seiberg, M.~Strassler, C.~Vafa and E.~Witten
for useful discussions. 
This research was conducted during the period S.G.
served as a Clay Mathematics Institute Long-Term Prize Fellow.
The work of S.G. is also supported in part by grant RFBR No. 01-02-17488,
and the Russian President's grant No. 00-15-99296.
D.T. is a supported by a Pappalardo fellowship and, in part, by funds 
provided by the U.S. Department of Energy 
(D.O.E.) under cooperative research agreement \#DF-FC02-94ER40818.


\end{document}